\documentclass[
    ,final            
  ]
  {aipproc}

\layoutstyle{6x9}

\begin{document}

\title{Klio: A 5 micron camera for the detection of giant exoplanets}

\author{Melanie Freed}{
  address={Steward Observatory, 933 N. Cherry Ave, Tucson, AZ 85721\\freed@as.arizona.edu}
}

\author{Philip M. Hinz}{
  address={Steward Observatory, 933 N. Cherry Ave, Tucson, AZ 85721\\freed@as.arizona.edu}
}

\author{Michael R. Meyer}{
  address={Steward Observatory, 933 N. Cherry Ave, Tucson, AZ 85721\\freed@as.arizona.edu}
}

\begin{abstract}

We plan to take advantage of the unprecedented combination of low thermal background and high resolution provided by the 6.5m MMT's adaptive secondary mirror, to target the 3-5 micron atmospheric window where giant planets are expected to be anomalously bright.  We are in the process of building a 3-5 micron coronograph that is predicted to be sensitive to planets as close as 0.4 arcsec to the parent star.  We expect to be able to detect giant planets down to 5 times Jupiter's mass for a 1 Gyr old system at 10 pc.  We plan to carry out a survey which is complementary to the radial velocity detections of planets and constructed to characterize the prevalence and distribution of giant planets around nearby, Sun-like stars.

\end{abstract}

\maketitle


\section{Why a 5 micron coronograph?}

With the number of known extrasolar planets now exceeding 100 (http://www.obspm.fr/ encycl/catalog.html), we have reached the point where we can begin to study the statistical properties of these planets (i.e. \cite{marcy}).  However, while radial velocity surveys have provided a wealth of information about exoplanets, they are unable to determine the inclination of the systems or any spectral information.  In addition, they are limited to relatively small star-planet separations over a practical baseline.  Other indirect methods such as astrometry and microlensing are also valuable tools for the detection of exoplanets.  However, only by direct detection can any spectral information about the exoplanets be determined, which is crucial to studying their physical properties (e.g. temperature, surface gravity, composition).  

So far, the transit method is the only direct technique that has successfully detected an exoplanet \cite{charbonneau} \cite{konacki}.  While transit techniques clearly have the potential to provide a wealth of information about exoplanets, they are also limited to close in planets, where multiple transits can quickly be observed and exoplanets are more likely to transit their primary star.   

In an effort to directly detect exoplanets at separations left largely unexplored by current measurements, we are building a coronographic imager to be used in the thermal infrared with ground-based adaptive optics (AO).  The camera will be optimized to target the thermal IR (3-5 microns), where exoplanets are thought to peak in brightness\cite{sudarsky}.  Until now, it has been impossible to use the thermal IR since no appropriate space-based telescope exists and the sky background is so large from the ground.  Additional emissivity introduced by conventional AO systems overwhelms any astronomical signal.

\section{The adaptive secondary at the MMT}

The adaptive secondary mirror at the 6.5m MMT provides a solution to the problem of increased emissivity from conventional adaptive optics systems.  By making the secondary mirror of the telescope itself the deformable surface, approximately eight additional surfaces are eliminated compared to typical AO systems \cite{lloyd-hart2000}.  This results in both an emissivity and throughput that are similar to a non-AO equipped telescope.  For faint objects, a conventional AO system at the MMT would need to integrate 2-3 times longer than the adaptive secondary to achieve the same signal-to-noise in L and M bands \cite{lloyd-hart2000}. 

The adaptive secondary mirror is now operational at the MMT telescope \cite{wildi} \cite{brusa}.  Measurements taken with the BLINC/MIRAC mid-IR camera give an emissivity of 6.5\% of the telescope plus adaptive optics system, which is typical of a telescope with no adaptive optics.  In January 2003, BLINC/MIRAC also recorded the first high order AO images taken in the thermal IR.  These were taken at 10.3 microns and show an achieve Strehl ratio (SR) of 0.96 with the AO system operational.  Once vibration-related issues are resolved, the adaptive secondary should be able to achieve SR=0.9 at 5 microns.  This number is based on actual observations at H band with the vibrations removed in post-processing.

\section{The 5 micron coronograph}

Klio will be a three channel system that consists of f/20, f/35, and pupil imaging modes.  The primary design driver for Klio is use in the high background regime (i.e. L and M bands).  The f/20 channel has been optimized for the L and M bands with Nyquist sampling at L band (0.048''/pixel) and a field of view of 15 x 12''.  The f/35 channel is optimized for the H and K bands with Nyquist sampling at K band (0.027''/pixel) and a FOV of 8.7 x 7''.  

\begin{figure}[h]
  \includegraphics[height=.253\textheight]{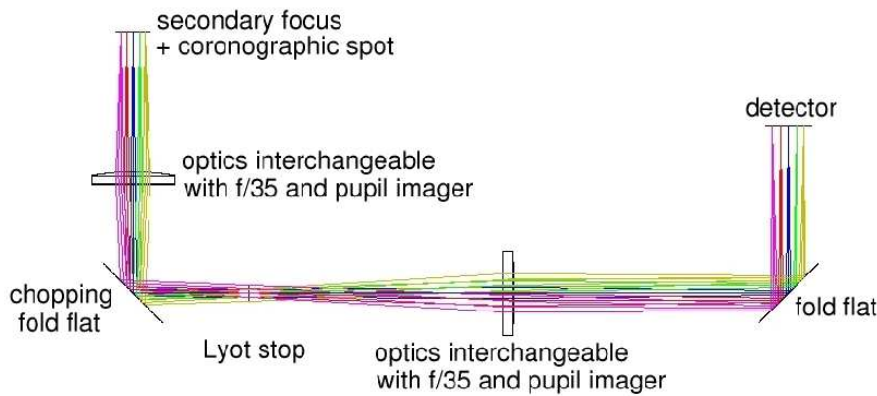}
  \caption{\label{Klio}Optical drawing of the f/20 channel of Klio.}
\end{figure}

Figure \ref{Klio} shows the optical design of the f/20 channel.  At the secondary focus of the telescope, a wheel will allow a selection of coronographic stops to be moved into the beam.  Optics will then form an image of the secondary, where another wheel will allow a selection of Lyot stops to be introduced into the beam before the final image is formed 

\noindent on the detector.  The detector is an Indigo 320x256 pixel InSb chip with 30 micron pixels.  We plan to have an engineering run without AO in the summer of 2004 and an engineering run with AO in the winter of 2004.

\section{Exoplanet survey}

So far, the only spectral information available for an exoplanet comes from observations of HD 209458b \cite{charbonneau2001}.  As a result, our understanding of exoplanets, especially those at Jupiter-like separations, would benefit greatly from even a single direct detection.  With ages derived from their primary stars, IR fluxes would provide anchor points for models that are essentially untested.  In addition, a carefully constructed survey has the ability to provide statistical information about the exoplanet population, even for a null result.

Radial velocity surveys provide a baseline from which we can make predictions about the distribution of exoplanets at further separations from their primaries.  Currently, results indicate a companion mass distribution of $dn/dM \sim M^{-0.7}$ and a separation distribution of $dn/dloga \sim a^{>0}$ \cite{marcy}.  A fit to the data (for 0.2 AU < a < 2 AU) gives $dn/da \sim a^{0.7}$ as a starting point.  Our survey will enable a quantitative test of the hypothesis that the mass and separation distribution of outer planets are consistent with those observed within 5 AU.  

As a first order prediction, we can simply extend the observed distributions to larger separations.  We performed a series of Monte Carlo simulations where the radial velocity mass distribution was taken to be valid from 1-15 $M_{Jupiter}$ and the separation distribution was extended from 0.01-50 AU.  An outer boundary of 50 AU was chosen to be consistent with the outer edge of our Kuiper Belt \cite{trujillo} as well as the radius of the smallest silhouette disk in Orion \cite{bally}.  Estimates of our sensitivities versus separation for L and M band are given in Heinze, Hinz, \& McCarthy (2003) \cite{heinze}.  Here we consider the 0.45'' gaussian

\begin{figure}[h]
  \includegraphics[height=.26\textheight]{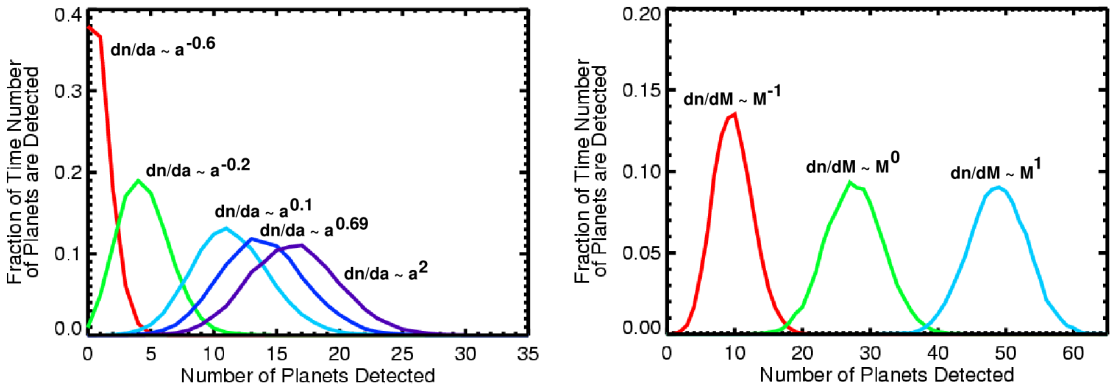}
  \caption{\label{carlo}(left) Fractional chance of detecting various numbers of planets given a mass distribution of $dn/dM \sim M^{-0.7}$ and 5 different separation distribution functions.  It appears unlikely that no planets will be detected with this survey.  If, indeed, no planets are detected, we should be able to significantly rule out a large parameter space of distribution functions.  (right) Fractional chance of detecting various numbers of planets given a separation distribution function of $dn/da \sim a^{0.7}$ and three different mass distribution functions.  All three mass distributions are clearly distringuishable.}
\end{figure}

\noindent coronograph with 2 hour integrations at the MMT.  

These simulations were used to define an optimal survey sample of 80 M0-F0 stars within 20 pc and less than 1 Gyr old.  With this sample, and assuming $dn/da \sim a^{0.7}$ and $dn/dM \sim M^{-0.7}$, we expect to detect 14 $\pm$ 3 companions with 5-15 $M_{Jupiter}$ at 13-50AU.  

We do not necessarily expect our simple extrapolation to be valid given that at very small separations planets may have very different evolutionary histories than their larger separations counterparts (e.g. migration \cite{Trilling}).  Our survey will provide an indication of the shape of the separation and mass distribution functions based on the number of companions detected.  

Figure \ref{carlo} indicates our chances of detecting various numbers of planets given $dn/da \sim a^{0.7}$ as well as three different mass distribution functions of the form $dn/dM \sim M^{\alpha}$ with $\alpha$=[-1,0,1].  These three distribution functions are clearly distinguishable for the given sample.  Assuming the given separation distribution function, we would be able to rule out $dn/dM \sim M^{-1}$ at 3$\sigma$ and $dn/dM \sim M^{1}$ at 11$\sigma$ in the event of no planet detections.  

The same process can be repeated for various separation distributions.  Figure 6 indicates our chances of detecting various numbers of planets, while this time keeping $dn/dM \sim M^{-0.7}$ and varying the separation distributon function from $dn/da \sim a^{\beta}$  $\beta$=[-0.6,-0.2,0.1,0.7,2].  For a null result we can rule out $\beta$=-0.2 at 2$\sigma$ and $\beta$=0.7 at 4$\sigma$.  

These limits will provide valuable constraints on theories of planet formation and evolution as well as help focus the next generation of direct detection surveys.

\begin{theacknowledgments}

We would like to thank IRLabs for their expert help, especially Elliott Solheid and Ken Salvestrini for designing the instrument dewar and testing the detector.  Thanks also to Michael Lesser for helping us integrate AZCam with Klio.  M.Freed acknowledges support from the NASA Graduate Student Researchers Program (NGT5-50394).  We also acknowledge support from AFOSR as well as the NASA Astrobiology Institute ``Laplace Center'' node at the UofA.

\end{theacknowledgments}

\bibliographystyle{aipproc}

\begin{thebibliography}
\expandafter\ifx\csname natexlab\endcsname\relax\def\natexlab#1{#1}\fi
\providecommand{\enquote}[1]{``#1''}
\expandafter\ifx\csname url\endcsname\relax
  \def\url#1{\texttt{#1}}\fi
\expandafter\ifx\csname urlprefix\endcsname\relax\def\urlprefix{URL }\fi

\bibitem[Marcy et al. (2003)]{marcy}
Marcy, G., Butler, R.P., Fischer, D.A., \& Vogt, S.S. 2003, in ASP Conf. Ser., Scientific Frontiers in Research on Extrasolar Planets, ed. D.Deming \& S. Seager (San Francisco:ASP)

\bibitem[Charbonneau et al. (2000)]{charbonneau}
Charbonneau, D., Brown, T.M., Latham, D.W., \& Mayor, M. 2000 ApJ, 529, L45.

\bibitem[Konacki et al. (2003)]{konacki}
Konacki, M., Torres, G., Jha, S., \& Sasselov, D.D. 2003, Nature, 421, 507.

\bibitem[Sudarsky, Burrows, \& Hubeny (2003)]{sudarsky}
Sudarsky, D., Burrows, A., \& Hubeny, I. 2003, ApJ, 588, 1121.

\bibitem[Lloyd-Hart (2000)]{lloyd-hart2000}
Lloyd-Hart, M. 2000, PASP, 112, 264.

\bibitem[Wildi, Brusa, \& Lloyd-Hart (2003)]{wildi}
Wildi, F.P., Brusa, G., \& Lloyd-Hart, M.  2003, Proc. SPIE, 5169

\bibitem[Brusa-Zappellini et al. (2003)]{brusa}
Brusa-Zappellini et al. 2003, Proc. SPIE, 5169

\bibitem[Charbonneau et al. (2001)]{charbonneau2001}
Charbonneau, D., Brown, T.M., Noyes, R.W. \& Gilliland, R.L. 2001, ApJ, 568, 377.

\bibitem[Trujillo \& Brown (2001)]{trujillo}
Trujillo, C.A. \& Brown, M.E.  2001, ApJ, 554, L95

\bibitem[Bally, O'Dell, \& McCaughrean (2000)]{bally}
Bally, J., O'Dell, C.R., \& McCaughrean, M.J.  2000, AJ, 119, 2919

\bibitem[Heinze, Hinz, and McCarthy (2003)]{heinze}
Heinze, Hinz, and McCarthy 2003, Proc. SPIE, 4839, 1154

\bibitem[Trilling, Lunine, \& Benz (2002)]{Trilling}
Trilling, D.E., Lunine, J.I., \& Benz, W. 2002, A\&A, 394, 241

\end{thebibliography}

\end{document}